\documentstyle[preprint,aps,eqsecnum,epsf,prd]{revtex}

\begin{document}

\draft
\preprint{LBL-38363}

\title{Chemical relaxation time of pions in hot hadronic matter}
\author{Chungsik Song and Volker Koch}
\address{ Nuclear Science Division, MS70A-3307,\\
          Lawrence Berkeley National Laboratory, Berkeley, CA 94720, USA}

\maketitle 

\begin{abstract}

We calculate characteristic time scales for chemical equilibration of pions 
in hot hadronic matter using an effective chiral Lagrangian. 
We find that inelastic
processes involving the vector and axial vector mesons 
reduce the chemical equilibration time by a factor of $\sim 10$ compared to the
result previously calculated in chiral perturbation theory. 
For a temperature of $T\sim 150 \, \rm MeV$ we obtain a
chemical relaxation time of $\tau_{ch} \simeq 10 \, \rm fm/c$, which is
comparable with typical time scales for a hadronic system generated in
SPS-energy heavy ion collisions. The effect of baryons is also estimated and
found to be negligible for SPS-energies but important for AGS-energies.
We predict, that chemical freeze-out should take place at considerable higher
temperatures $\Delta T \simeq 20 \, \rm MeV$ than thermal freeze-out and that 
the hadronic phase would not sustain a pion chemical potential larger 
than $100 \, \rm MeV$.
\end{abstract}
\bigskip


\section {Introduction}

High energy nucleus-nucleus collisions offer a unique opportunity to explore
the large-scale properties of quantum chromodynamics, in particular its phase
structure at high temperatures and densities \cite{rhic}.  
For sufficiently high collision energy,
we expect a phase transition into a locally thermalized, 
deconfined plasma of quarks and gluons, the so called quark-gluon plasma 
(QGP) \cite{qgp}. 
There may be two different phase transitions,
deconfinement and chiral phase transition, in QCD at high temperature and/or
density. The nature and order of
the phase transition have been studied as well as possible signatures for
the new phase of hadronic matter.

Certainly the system produced in these collisions is not immediately in 
thermal and chemical equilibrium. 
Secondary interactions among the produced particles are necessary to achieve
equilibrium. If equilibrium is reached, 
global observables such as transverse energy
production can be related to thermodynamic variables, such as energy and
entropy density, commonly used to characterize these collisions. 
One can also use the hydrodynamic equations to study the evolution of 
the hadronic system at finite temperature and/or density. 
These conditions considerably facilitate the calculation  of certain
quark-gluon plasma signals, such as the yields of photons, 
lepton pairs, strangeness and hadrons containing heavy quarks.
However, even if the heavy ion collision is energetic enough to 
fully equilibrate the
system at early time, e.g. in the quark-gluon plasma phase, the system will not
necessarily remain equilibrated as it expands.
Only when the collision rate is greater than the expansion rate
will equilibrium be maintained. 
It is therefore crucial to understand the following questions:
How does the system approach equilibrium? 
What are the time scales for thermal and chemical equilibration?,
and to which extent  is equilibrium maintained in the hadronic phase before 
the system breaks up into the final state hadrons?

Recently the question of thermodynamic equilibration in the early stages 
of collisions at RHIC and LHC have been studied in the parton model 
\cite{shuryak,wang,geiger}.  
In the partonic system created early in the collisions, gluons are 
the dominant components of the system. They reach thermal equilibrium on the
time scale $0.3\sim 0.5\>\rm fm/c$ after collision with an undersaturated gluon
density. 
The thermalization of quarks follows that of gluons with the time scale $1\>\rm
fm/c$ and reach  chemical
equilibrium later since the quark pair production by gluon fusion and the decay
of gluons occur at a somewhat smaller rate.
These thermodynamic equilibration conditions would be observed in open charm, 
photon, dilepton, and $J/\psi$ production.

The question of the equilibration in hadronic phase, on the other hand, 
has been addressed by analyzing the observed hadron abundances 
and spectral distributions \cite{BM,BM_AGS,BM_SPS,rafelski,cleymans,heinz}.
The prevailing view requires  first a chemical 
particle freeze-out, from which particle abundances are preserved, 
followed by a thermal freeze-out, after which all interactions between 
the produced particles have ceased.   
Since the system remains in local thermal equilibrium
the state can be characterized by a maximum
of the entropy consistent with the conservation laws for energy, 
momentum and of
the relevant particle number. Neglecting dissipative processes, the
corresponding continuity equations then imply that entropy is conserved, 
i.e. the expansion is an adiabatic process. 
If so, the hadronic system is described by two parameters, temperature
$T(x)$ and a `chemical potential'\footnote{A detailed discussion of the meaning
of the `chemical potential' will be given at the beginning of section
\ref{chem_equ}.} $\mu_i(x)$ for each species. 
Assuming such a  partial equilibrium, i.e. thermal but not chemical,  
the observed hadronic abundance  can be well described.

As far as strange degrees of freedom are concerned, one expects only a partial
saturation since the time scale for strangeness saturation is
normally larger than the collision time scale.
The chemical properties of strange particle at freeze-out are regarded as an
important ingredient to understand the strangeness production in high energy
nucleus-nucleus collisions \cite{strange}. 

Naturally, hadronic observables can only tell us something about the conditions
at freeze-out. 
In this paper, we are interested in the thermal and chemical equilibrium
condition of pions in hot hadronic matter at temperatures  higher than the 
freeze-out temperature in order to see if the observed freeze-out conditions
can be understood within a pure hadronic model.
We use an effective chiral Lagrangian with vector and axial
vector mesons to describe the interaction of pions in hadronic matter. 
In section 2, we first consider the processes that lead to 
thermal equilibrium for pions in hot hadronic matter. 
Elastic pion collisions, $\pi+\pi \to \pi+\pi$, turn out  to be the 
principal thermalizing process.
Secondly, we clarify the meaning of chemical equilibrium and chemical potential
of pions in hot matter produced in high energy nucleus-nucleus collisions. 
Thirdly, we estimate the characteristic time scale for 
chemical equilibrium of pions in hadronic matter, first with pions, and then
later including  resonances.  Finally, we also explore the effect of baryons on
the chemical equilibrium conditions of pions in hot hadronic matter. 
In section 3, the time scales for the chemical equilibrium of pions 
are studied assuming a finite pion chemical potential. 

\section {Thermodynamic equilibration of pions}

The thermodynamic equilibration in a hadronic system is driven by  
multiple collisions among particles in the system.
Generally the  equilibration time is directly proportional to the collision
rate. In an expanding system, therefore, the collision rate 
should be much larger than the expansion rate of the system in order to 
maintain thermodynamic equilibrium. As long as the expansion
velocity of the system does not exceed the most probable velocity of 
the thermal particles in the system, this condition is satisfied 
if the mean free path of a particle
is shorter than the size of the system. 
In the following we determine  the thermal and chemical 
relaxation time of pions  assuming that the 
system is slightly deviated from the equilibrium state, i.e. in the relaxation
time approximation.

\subsection{Thermal equilibrium}

At low temperatures hadronic matter, 
which mainly consists of pions,
can be analyzed systematically in the framework of chiral perturbation
theory at finite temperature. 
The low energy theorems, which are obtained from
current algebra and partial conservation of axial-vector current (PCAC), can be
translated into a corresponding set of exact statements concerning the
coefficients of the low temperature expansion. 
Since typical pion energies are of the order $E\sim T$ at finite $T$, 
interactions among pions generate power corrections, 
controlled by the expansion parameter
$T^2/f_\pi^2$, where $f_\pi=93$ MeV is the pion decay constant \cite{chpt}. 
It is therefore possible to treat the interaction in low temperature 
pion gas as a perturbation.

The thermal equilibration of pions at low temperatures 
will be governed by elastic two-body collisions, $\pi+\pi\to\pi+\pi$,
since inelastic processes contribute 
in the next leading order  in the low energy expansion. 
As long as we consider elastic two-body collisions, 
the thermal relaxation time $\tau_{th}$ is given by \cite{thermal}
\begin{eqnarray}
{1\over\tau_{th}} &=&{1\over4}(1-e^{-\beta E_a}){1\over 2E_a}
    \int {d^3p_2\over(2\pi)^3 2E_2}
         {d^3p_3\over(2\pi)^3 2E_3}
         {d^3p_4\over(2\pi)^3 2E_4}
\nonumber\\[2pt]
&&\qquad\qquad\qquad\qquad
\times (2\pi)^4\delta^4(p_a+p_2-p_3-p_4)
\nonumber\\[2pt]
&&\qquad\qquad\qquad\qquad
\times\sum_{2,3,4}
\vert {\cal M}(\pi_a\pi_2\to\pi_3\pi_4)\vert^2 f_2(1+f_3)(1+f_4),
\end{eqnarray}
where $f_i$'s are Bose-Einstein distribution function.
The factor $1/4$ in front comes from identity of particles at initial and 
final state and
the sum is over the spin and isospin degeneracy of particles $2,3$ and $4$. 
The second term ($\sim e^{- \beta E}$) 
indicates the contribution from inverse reaction.
Note that this result is obtained in the relaxation time approximation.
In the classical limit where Bose-Einstein or Fermi-Dirac quantum statistics is
approximated by the Boltzmann statistics,  this can be written as 
\begin{equation}
{1\over\tau_{th}}={1\over 2}(1-e^{-\beta E_a})
    \int {d^3p_2\over(2\pi)^3}\sum_2\sigma(a2\to 34)v_{a2}e^{-E_2/T},
\end{equation}
where $\sigma$ is the cross section corresponding to the relevant reaction and
$v_{a2}$ is the relativistic velocity.

The mean relaxation time is defined as 
\begin{equation}
\bar\tau_{th}={1\over \int d^3p_a f_a(p_a)}\int d^3p_a\tau_{th}(p_a)f_a(p_a).
\end{equation}
When we use the current algebra result for the $\pi\pi$ scattering 
cross section we obtain 
\begin{equation}
\bar\tau_{th}\approx {12 f_\pi^4\over T^5}.
\label{tau_thermal}
\end{equation}
With this approximation it has been shown that the mean relaxation
time is about 2 fm/c at $T=150$ MeV \cite{thermal}. 
Since the mean free path $\lambda\approx\tau$ in the relativistic limit, the
result can be compared to the size of the system.
Here we assume that the radius of the hot matter
would be 5 $\sim$ 10 fm.
This implies that pions are in thermal equilibrium in hadronic matter,
mainly due to the two body elastic collisions. 

At temperatures close to the phase transition, 
other heavy mesons like kaons, vector mesons, etc., 
become increasingly abundant and reduce the thermalization time even further.
Thus, we conclude that pions can maintain thermal
equilibrium in hot hadronic matter even at comparatively low temperatures.
From formula Eq.~(\ref{tau_thermal}) 
we expect a freeze-out temperature $\leq 130 \,
\rm MeV$ for a small system such as $\rm S+S$ or $\rm S+Au$. This is somewhat
lower than the values extracted from experiment, which are $\sim 150 \, \rm
MeV$. The reason for this slight discrepancy is most likely 
due to the flow generated in the reaction, which reduces 
the effective system size.

We also expect that light vector mesons, $\rho,\>\omega,\>\phi$, reach 
thermal equilibrium in hot hadronic matter. 
For $\rho,\>\omega$ the dominant reactions will be the collisions with thermal
pions through heavy resonances such as $\pi+\rho\to a_1\to\pi+\rho$ 
and $\pi+\omega\to b_1\to\pi+\omega$ , see Ref. \cite{coll}.
In this reference it  has also been shown that the various collisions in hot 
hadronic matter make it possible for phi mesons to be in thermal equilibrium.

\subsection{Chemical potential of pions in hot hadronic matter}
\label{chem_equ}

Before we embark on the discussion of chemical equilibrium we should first
clarify what we mean when we talk about a `chemical potential' in the
subsequent paragraphs. Strictly speaking, a chemical potential is 
associated with a conserved quantity. In case of pions, the conserved quantity
would be the charge and, therefore, the only rigorously defined 
chemical potential would be associated with the charge of the pions (and all
other particles involved in the system for that matter). This chemical
potential, let's call it $\mu_{charge}$, enters with opposite signs in the
Bose-Einstein distribution functions for $\pi^+$ and $\pi^-$. In the following
discussion we will always consider a charge neutral system and, therefore, 
$\mu_{charge} = 0$. 

However, the lifetime of a system created in a relativistic heavy ion
collisions is much too short in order for the electromagnetic and weak
processes to be relevant, so that we can safely ignore them from now on. 
Let us also 
assume for a moment, that the time scale for the strong number changing 
processes are long compared to the lifetime of the system, whereas the
elastic 
processes, which are responsible for the kinetic equilibration, are fast. In
this case, the number of pions does not change, and, after kinetic equilibrium
has been established, the pions are distributed according to a Bose-Einstein
distribution 
\begin{equation}
n(E) = \frac{1}{e^{(E - \mu_f)/T} - 1},
\end{equation}
where $\mu_f$ is now the chemical potential associated with the conserved pion
number. Note, that $\mu_f$ enters with the same sign for all charge states of
the pions. In the literature, the factor $e^{\mu_f / T}$ is often called a
fugacity. Of course in the true thermodynamic limit where the system has
an infinite amount of time to change the number of particles, $\mu_f = 0$. 
Therefore, a non vanishing fugacity parameter, $\mu_f \neq 0$, is often 
referred to an indication of chemical non-equilibrium. This concept can lead
to some confusion, as we shall outline now.

First of all, let us stress the difference between what we shall call {\em
true} number changing processes, such as 
\begin{equation}
\pi \pi \rightleftharpoons \pi \pi \pi \pi
\end{equation}
and {\em apparent} number changing processes such as
\begin{equation}
\pi \pi \rightleftharpoons \rho.
\end{equation}
As  we will demonstrate in the following, 
the first process drives the fugacity
parameter $\mu_f$ to zero, leading to  an {\it absolute} chemical equilibrium.
In the second process, on the other hand, 
the pions are simply `locked up' in the
$\rho$-meson, but never really disappear. This process ceases to be effective
once $2 \mu_f^\pi = \mu_f^\rho$ with $\mu_f^\pi$ no necessarily vanishing. 
Thus, pions and $\rho$-mesons can be chemically equilibrated with each other at
any value of $\mu_f^\pi$. This condition we will refer to as {\em relative}
chemical equilibrium.
Consequently, this process and many other of that
kind, do not drive the system to {\it absolute} chemical equilibrium, 
i.e. $\mu_f^\pi = 0$.

Whenever in the following discussion  we refer to a `chemical potential' we
actually talk about the fugacity parameter, $\mu_f$. Whenever we talk about
chemical equilibrium, we mean the {\em absolute} chemical equilibrium. 
When particles
such as the $\rho$-meson and the pions are 
in chemical equilibrium with respect to
each other, we will talk about a {\em relative} chemical equilibrium.
After these hopefully clarifying remarks let us now turn to the calculation
of the chemical relaxation times.

\subsection{Chemical relaxation time for pions}

The characteristic time associated with chemical equilibrium is determined by
a large number and variety of inelastic processes.  
The following processes are subset of possible reactions in hadronic
matter which change the number of pions;
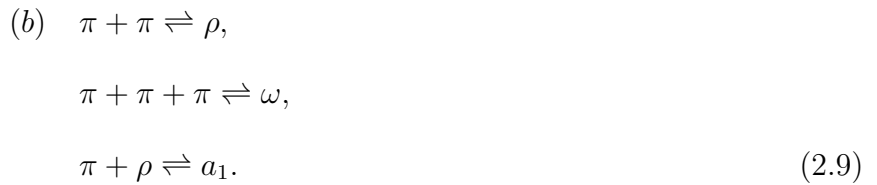
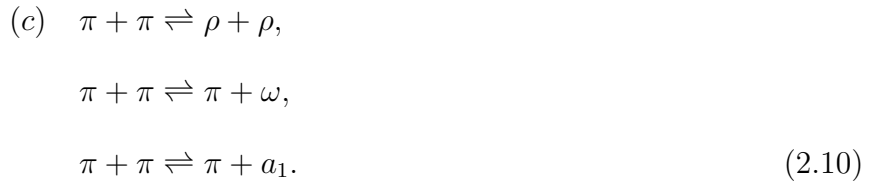
\begin{eqnarray}
&(a)&\quad  \pi+\pi\rightleftharpoons\pi+\pi+\pi+\pi.
\label{in-true1}\\[12pt]
&(b)& \quad \pi+\pi \rightleftharpoons\rho,\nonumber \\
&&\quad  \pi+\pi+\pi\rightleftharpoons\omega,\nonumber \\
&&\quad  \pi+\rho\rightleftharpoons a_1.
\label{in-apparent}\\[12pt]
&(c)&\quad  \pi+\pi \rightleftharpoons\rho+\rho, \nonumber\\
&&\quad  \pi+\pi \rightleftharpoons \pi+\omega,\nonumber\\
&&\quad  \pi+\pi \rightleftharpoons \pi+a_1.
\label{in-true2}
\end{eqnarray}

In order to estimate the characteristic time scales for chemical equilibrium 
we consider a system slightly out of equilibrium.
For this case the distribution function in phase space
of the various particles takes the form 
\begin{equation}
f(p,x)={1\over \exp[\beta_\nu(p^\nu-\lambda^\nu)]-1},
\end{equation}
where $p_\nu$ is the four-momentum of the particle. $\lambda_\nu(x)$ and
$\beta_\nu(x)$ are four vectors constructed from the local flow velocity
$u_\nu(x)$, the local chemical potential $\mu(x)$ and the inverse local
temperature $\beta(x)=1/T(x)$:
\begin{eqnarray}
\lambda_\nu(x) &=& \mu(x) u_\nu(x),\\
\beta_\nu(x) &=& \beta(x) u_\nu(x).
\end{eqnarray}
As explained in the beginning, the non-vanishing chemical potential 
$\mu$ is the parameter which
measures to which extent the system deviates from {\em absolute} chemical 
equilibrium. 

For simplicity we consider hadronic
matter in which all thermodynamic variables depend only on time and are
uniform in space. The change in particle density ($n$) is then 
given by Boltzmann type rate  equations \cite{matsui};
\begin{equation}
{d n(t)\over dt}=\delta n\biggl(R_{\rm gain}-R_{\rm loss}\biggr),
\label{par_number}
\end{equation}
where $\delta n$ is the number of particles changed in the process and 
$R_{\rm gain}$,  $R_{\rm loss}$ are standard Boltzmann collision 
integrals
for the production and annihilation rates respectively.

\subsubsection{Pion gas}

To be specific, let us consider first hadronic matter at low temperatures which
mainly consists of pions. Since chiral symmetry suppresses the reaction 
rates for multi-pion
scattering we consider an inelastic reaction with the minimum number of pions,
i.e. the reaction $\pi+\pi\rightleftharpoons\pi+\pi+\pi+\pi$.
In this case the gain term, i.e. the term that {\em increases} the pion number
is associated with the reaction $\pi+\pi \rightarrow \pi+\pi+\pi+\pi$ 
and the loss term is related with the backwards process.
Using the unitarity relation for the matrix element 
${\cal M}$
\begin{equation}
\vert {\cal M}(\pi\pi \to \pi\pi\pi\pi)\vert^2
      =\vert {\cal M}(\pi\pi\pi\pi \to \pi\pi)\vert^2,
\end{equation}
the rate of change for the pion number is given by
\begin{eqnarray}
{d n_\pi(t)\over dt} &=& 2{\cal S}\,
\int{d^3p_1\over (2\pi)^3 2E_1}\int{d^3p_2\over (2\pi)^3 2E_2}
\cdots\int{d^3p_6 \over (2\pi)^3 2E_6}
\nonumber\\[4pt] 
&&
\times \,\, (2\pi)^4\delta^{(4)}(p_1 + p_2 - p_{3} - \cdots-p_6)
\nonumber\\[4pt] 
&&
\times \sum_{1,2,\ldots,5,6}\vert{\cal M}
(\pi_1\pi_2 \rightarrow \pi_3\cdots\pi_6 )\vert^2  
\nonumber\\[4pt]
&&
\times \,\, \left[ f_{1}(p_{1})\cdots f_2(p_2) \,
(1+f_3(p_3)) \cdots (1+f_6(p_6)) \right. 
\nonumber \\
&& \,\,\,\,\ 
- \left. f_3(p_3) \cdots f_6(p_6) \, (1+f_1(p_1))(1+f_2(p_2)) \right].
\end{eqnarray}
Here, the pion density $n_\pi(t)$ is the sum over all charge states. 
The statistical factor ${\cal S}$ for 
the initial and final state is given by  ${1\over 2!\,4!}$ and 
the sum is over the isospin degeneracy of all participating particles.
In the  Maxwell-Boltzmann limit of the
Bose-Einstein quantum statistics, we have 
\begin{equation}
{dn_\pi\over dt}=
2\,(e^{2 \mu_\pi/T}-e^{4 \mu_\pi/T})I_0(T),
\label{four_pi}
\end{equation}
where
\begin{eqnarray}
I_0(T)&=& {\cal S} \,
\int{d^3p_1\over (2\pi)^3 2E_1}\int{d^3p_2\over (2\pi)^3 2E_2}
\cdots\int{d^3p_6 \over (2\pi)^3 2E_6}
\nonumber\\[4pt] 
&&
\times \,\, (2\pi)^4\delta^{(4)}(p_1 + p_2 - p_{3} - \cdots-p_6)
\nonumber\\[4pt] 
&&
\times \sum_{1,2,\ldots,5,6}\vert{\cal M}
(\pi_1\pi_2 \rightarrow \pi_3\cdots\pi_6 )
\vert^2 e^{-(E_1+E_2)/T}. 
\end{eqnarray}

Notice, that the number of pions ceases to change if and only if 
$\mu_\pi = 0$. 
Thus, the process $\pi+\pi\rightleftharpoons\pi+\pi+\pi+\pi$ drives 
the system to
chemical equilibrium and, therefore, is a true number changing process
according to our previous definition. 
If we assume that the system is only slightly away from chemical equilibrium,
i.e. $\frac{\mu_\pi}{T} \ll 1$, we may expand the exponentials to first
order in $\frac{\mu_\pi}{T}$. Also, in this approximation, for fixed
temperature  
\begin{equation}
n_\pi(t) = n^0_\pi \left(1 + \frac{\mu_\pi(t)}{T}\right),
\end{equation}
where $n^0_\pi$ stands for the equilibrium density.
The above rate equation Eq.~(\ref{four_pi}) then reduces to
\begin{equation}
n^0_\pi \frac{d \mu_\pi}{d t} = - 4 \mu_\pi I_0(T).
\end{equation}
The solution is given by 
\begin{equation}
\mu_\pi(t) = \mu_\pi(0) \, \exp(-t/\tau_{ch}), 
\end{equation}
with 
\begin{equation}
{1\over\tau_{ch}} = \frac{4 I_0(T)}{n^0_\pi}.
\label{chem_1}
\end{equation}
The above approximations are nothing else but the well known relaxation time
approximation and  $\tau_{ch}$ is called the chemical relaxation time.

For the process under  consideration here the integral $I_0(T)$ is given by
\begin{equation}
I_0(T)= {1\over2}\int{d^3p_1\over (2\pi)^3}\int{d^3p_2\over (2\pi)^3} 
\sum_{1,2} \sigma(\pi_1\pi_2\to\pi_3\pi_4\pi_5\pi_6)v_{12} 
e^{-\beta E_1-\beta E_2},
\end{equation}
where $\sigma(\pi_1\pi_2\to\pi_3\pi_4\pi_5\pi_6)$ stands for the isospin
averaged cross section and $v_{12}$ for the relative velocity. 
The sum is over isospin of the initial states.
In the chiral limit the inelastic cross section is  \cite{goity}
\begin{equation}
\sum_{1,2}\sigma(\pi_1\pi_2\to\pi_3\pi_4\pi_5\pi_6)
={67\over 2^{17} 3^4 \pi^5}{s^3\over f_\pi^8}.
\end{equation}
With this cross section the chemical relaxation time is 
proportional to $f_\pi^8/T^9$ and
estimated to be about $26$ fm/c at $T=180$ MeV 
(about 40 fm/c with $m_\pi$=138 MeV at the same temperature).
This is very large compared to the size of the hot matter produced in
nucleus-nucleus collisions.
This implies that the chemical equilibrium in the 
reaction $\pi\pi\rightleftharpoons\pi\pi\pi\pi$ would not be reached 
in the expanding hot hadronic matter consisting of pions only. 

\subsubsection{Resonance Gas}

In the presence of excited states, i.e. resonances, 
the number of pions is changed 
by the decay process of resonances (and vice versa), e.g.,
$\rho\rightleftharpoons\pi\pi$, $\omega\rightleftharpoons\pi\pi\pi$,
$a_1\rightleftharpoons\pi\rho$,
and inelastic scattering such as $ \rho\rho\rightleftharpoons\pi\pi$, 
$\pi\omega\rightleftharpoons\pi\pi$, $\pi a_1\rightleftharpoons\pi\pi$ etc.

{\it (i) Resonance decays}

\noindent
First we consider resonance decays. 
In this case the relaxation time will be inversely proportional to the decay
width. Specifically, in case of $\rho\rightleftharpoons\pi\pi$ we have
\begin{equation}
\frac{d n_\pi(t)}{d t} = 2\,(e^{\mu_\rho/T} - e^{2 \mu_\pi/T})\,
I_0(\rho\rightleftharpoons\pi\pi;T). 
\label{rho_pipi}
\end{equation}
$I_0(\rho\rightleftharpoons\pi\pi;T)$ can be related to the decay width
of the $\rho$ via
\begin{eqnarray}
I_0(T,\rho\rightleftharpoons\pi\pi) &=& \sum_{\rho}\, 
\int{d^3p_\rho \over (2\pi)^3}\frac{E_\rho}{m_\rho}\Gamma(p_\rho)f_\rho(p_\rho)
\nonumber \\
&\approx& \, \bar\Gamma \, n_\rho(T),
\end{eqnarray}
where the sum is over spin and isospin of vector mesons. 
$\bar\Gamma_\rho$ is the decay width of neutral vector meson and 
$n_\rho$ is the sum over all spin and isospin states. 

Notice, that from Eq.~(\ref{rho_pipi}), the process
$\rho\rightleftharpoons\pi\pi$ ceases to change the number of pions once
$\mu_\rho = 2 \mu_\pi$. This may very well happen at a finite value $\mu_\pi$.
Consequently this process does not drive the system into {\em absolute} 
chemical equilibrium,
as we have defined it. It rather ensures a {\em relative} chemical equilibrium
between $\rho$-mesons and pions, which is characterized by the condition
$\mu_\rho = 2 \mu_\pi$. Following our definitions, the process 
$\rho\rightleftharpoons\pi\pi$ is an apparent number changing one, since it
only shuffles pions back and forth from the $\rho$-resonance. The fact that 
the pion chemical potential can take on a finite value means nothing else than
the conservation of the pion number in this type of process.

Expanding the distribution functions around the
{\em relative} chemical equilibrium , one can again derive a relaxation time,
$\tau^{rel}_{ch}$, which is a measure of how fast the {\em relative} chemical
equilibrium between pions and $\rho$-mesons is reached. Following the steps of
the previous section one finds
\begin{equation}
\tau^{rel}_{ch} = \frac{1}{2\bar\Gamma}
\left({n_\pi^0\over n_\pi^0+4 n_\rho^0}\right)= 0.4\sim 0.6 \> \rm fm/c,
\end{equation}
where $n_\alpha^0$ indicates the equilibrium density 
for each species. 
This is very short, and to a good approximation we can assume that pions and
$\rho$-meson are always in {\em relative} chemical equilibrium.  

The same conclusion holds for the decay $a_1\to\pi\rho$
since $\Gamma_{a_1\to\rho\pi}=400\>\rm MeV$. 
Thus we have $\mu_{a_1}=\mu_\pi+\mu_\rho=3\mu_\pi$ 
where we use $\mu_\rho=2\mu_\pi$.
However, the decay rate for omega mesons into three pions
is too small in order to reach {\em relative} chemical equilibrium.

{\it (ii) Inelastic scattering}

\noindent
Next we consider the inelastic collisions involving resonances such as, 
$\pi+\pi\rightleftharpoons\rho+\rho$,
$\pi+\omega\rightleftharpoons\pi+\pi$, and
$\pi+a_1\rightleftharpoons\pi+\pi$. 
Again, the rate of change of pion density is given by (here in case of 
$\pi+\pi\rightleftharpoons\rho+\rho$)
\begin{equation}
{dn_\pi\over dt}=
2\,(e^{2 \mu_\rho/T}-e^{2 \mu_\pi/T}) \, 
I_0(\pi+\pi\rightleftharpoons\rho+\rho;T).
\label{rhorho_pipi}
\end{equation}
As long as the process $\rho\rightleftharpoons\pi\pi$ is fast compared 
to the one we are
considering here, we can, to a good approximation, 
assume pions and $\rho$-mesons
are in {\em relative} 
chemical equilibrium, i.e. $\mu_\rho = 2 \mu_\pi$. With this
replacement, the above rate equation Eq.~(\ref{rhorho_pipi})
becomes similar to 
Eq.~(\ref{four_pi}) and, hence, the chemical relaxation time is given by 
(see Eq.~(\ref{chem_1}))
\begin{equation}
{1\over\tau_{ch}} = \frac{4 I_0(\pi+\pi\rightleftharpoons\rho+\rho;T)}
                    {n^0_\pi}.
\label{chem_2}
\end{equation}
The same arguments hold for the process $\pi+a_1\rightleftharpoons\pi+\pi$,
whereas the approximation will be not valid in case of 
$\pi+\omega\rightleftharpoons\pi+\pi$. For the reactions involving $\omega$
mesons we assume $\mu_\omega\approx 0$ to estimate the corresponding
contribution to the chemical relaxation time of pions. 

In order to evaluate the relevant collision integrals, hadronic cross sections
have been calculated from an effective chiral
Lagrangian with vector and axial vector mesons. The details are given in 
appendix A. All processes included in the calculation are summarized and 
results for invariant amplitudes are presented in Appendix B.
To take into account the off-energy-shell effects we introduce form 
factors in the calculation of the cross section.
The standard way to accomplish this is to insert a monopole form factor 
at each vertex in a $t-$channel diagram
\begin{equation}
F_\alpha={\Lambda^2-m_\alpha^2\over \Lambda^2-t},
\end{equation}
where $\alpha$ indicates a species for
the exchanged particle. We take a value of $\Lambda=1.7$ GeV for
$\rho\pi\pi$ vertices in $\rho$ exchange reactions \cite{formfactor}
and of $m_\rho$ in pion exchange processes. $\Lambda=m_{a_1}$ and
$1.7\,\rm GeV$ are taken for $a_1\pi\rho$ vertices and 
$\omega\rho\pi$ vertices, respectively. 

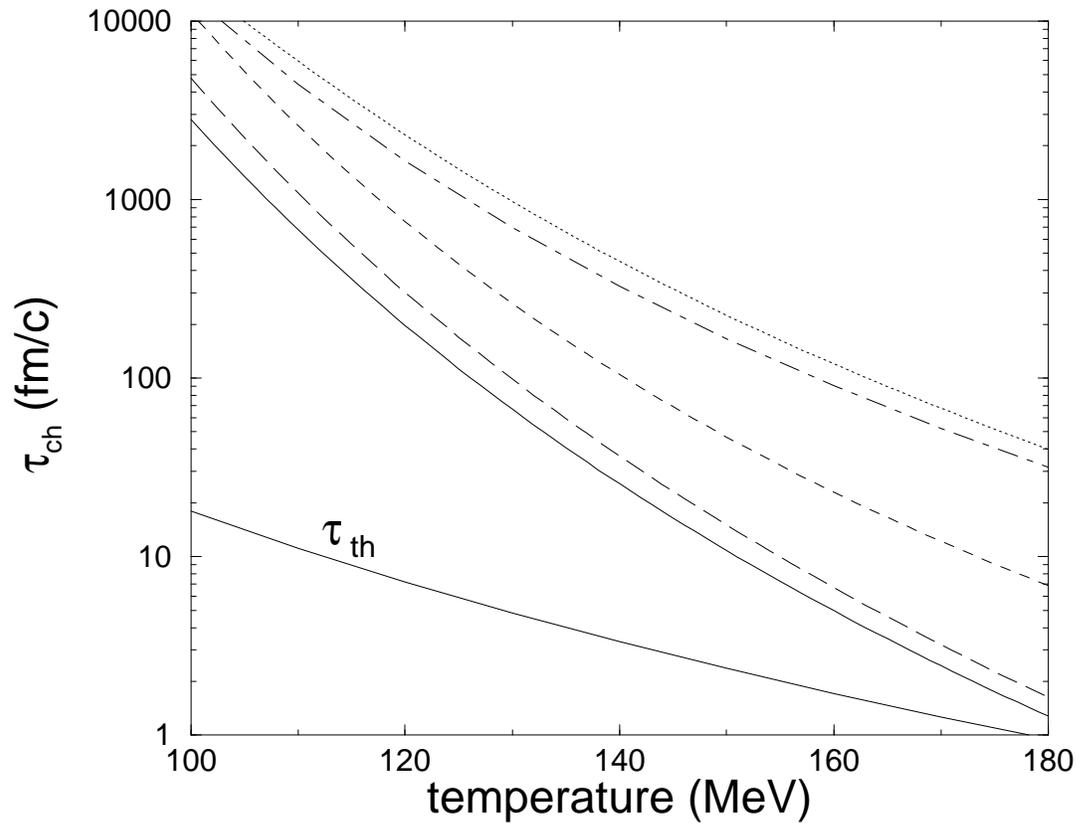
\begin{figure}
\caption{Chemical relaxation time of pions in hot hadronic matter.
The reactions $\rho\rho\leftrightarrow\pi\pi$ (long dashed curve),
              $\pi\omega\leftrightarrow\pi\pi$ (dashed-dot curve),
              $\pi a_1\leftrightarrow\pi\pi$ (dashed curve) and 
              $\pi\pi\pi\pi\leftrightarrow\pi\pi$ (dotted curve) 
              are considered. 
              The solid curve represents the total relaxation time. 
              These results are
              compared with the thermal relaxation time, $\tau_{th}$, obtained
              from Eq.~(2.4).
\label{ch_relax}}
\end{figure}

In Fig.~\ref{ch_relax} 
we show the chemical relaxation times for each processes 
as a function of temperature. 
For comparison the chemical relaxation time corresponding to the reaction  
$\pi\pi\pi\pi\rightleftharpoons\pi\pi$ is also shown \cite{goity}.
The total relaxation time is determined by including 
all true pion number changing processes,
with the exception of $\pi+\omega\rightleftharpoons\pi+\pi$, which is slow
because of the long lifetime of the $\omega$.
The result is given by the solid curve. Due to the hadronic resonances, 
the resulting relaxation time is almost
an order of magnitude shorter than that previously 
obtained within chiral perturbation theory \cite{goity}. 
We find that the relaxation time is about 1.5 fm/c at $T\sim 180 \,\rm MeV$
and increases to about 10 fm/c at $T=150$ MeV.

At a temperature of $T \simeq 150 \, \rm MeV$ the total relaxation time is 
about $10 \, \rm fm/c$, which is comparable to typical system sizes created in
ultrarelativistic heavy ion collisions involving heavy nuclei such as lead.
However, the corresponding temperature, for which the thermal relaxation 
time assumes the same value of $\tau_{th} = 10 \, \rm fm/c$ is considerably
low, namely about $110 \, \rm MeV$. From these numbers we expect, that even
with the inclusion of the resonance chemical freeze-out takes place before the
thermal one. 
Moreover, the above estimate has assumed that the resonances are
formed instantaneously. Thus, the chemical relaxation time, may be slightly
larger once the formation of the resonances is taken into account properly.

\subsubsection{Absorption on baryons}

Next, we take into account the pion number changing processes involving 
baryons.
The dominant contributions come from low laying baryons, 
like nucleon $N(938)$ and $\Delta(1236)$. 
Since the $\Delta$ resonance decays into $N\pi$ with width
$\Gamma_{\Delta N\pi}=120\> \rm MeV$ we expect there will be 
a {\em relative} chemical equilibrium 
with respect to the reaction 
$\Delta\to N\pi$ with $\mu_\Delta=\mu_B+\mu_\pi$.
Here $\mu_B$ indicates the baryon chemical potential, associated with the
conservation of the baryon number.

There are also inelastic, i.e true pion number changing,  reactions including 
baryon resonances, such as $NN\to N\Delta$. To estimate the 
characteristic time scale for the baryon
induced inelastic collisions we use a phenomenological parameterization for the
isospin averaged $N\Delta\to NN$ cross section \cite{delta};
\begin{equation}
\sigma({N\Delta\to NN})=(p_f^2/p_i^2){1\over 8}\sigma(\sqrt{s}),
\end{equation}
where $p_f$ is the momentum in the final $NN$ channel and 
\begin{equation}
\sigma(\sqrt{s})={20\times (\sqrt{s}-2.015)^2\over 0.015+(\sqrt{s}-2.015)^2}.
\end{equation}
Here $\sqrt{s}$ is in GeV, $\sigma$ in mb.
Then the time scale $\tau_{ch}$ is given by
\begin{equation}
{1\over \tau_{ch}}=-{1\over n_0}e^{2\mu_B/T}I_B(T),
\end{equation}
where we use $\mu_\Delta=\mu_B+\mu_\pi$ and
\begin{equation}
I_B(T)=\int{d^3p_N\over (2\pi)^3}\int{d^3p_\Delta\over (2\pi)^3} 
\sum_{1,2}\sigma({N\Delta\to NN})v_{12} 
e^{-\beta E_N-\beta E_\Delta}. 
\end{equation}
Here the sum is over the spin and isospin for the initial particles.

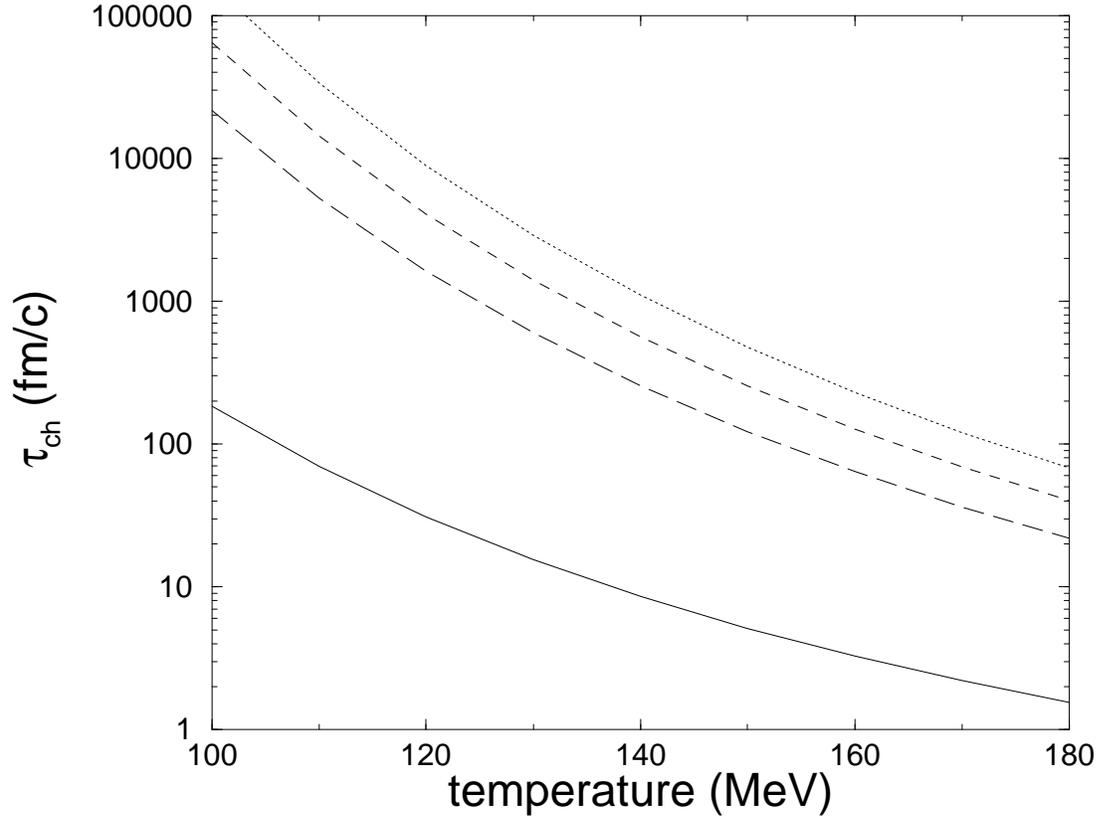
\begin{figure}
\caption{Contribution of baryon involved reactions to the chemical relaxation 
time of pions in hot hadronic matter.
Results are obtained with $\mu_B=176$ MeV (dotted line : SPS),
223 MeV (dashed line : SPS),
278 MeV (long-dashed line : SPS) and 
$\mu_B=$516 MeV (solid line : AGS).}
\label{baryon}
\end{figure}

Results are shown in Fig. \ref{baryon}. We use 
$\mu_B=176,\> 223,\> 278, \>516 \>\rm MeV$ for the baryon chemical 
potential which are
obtained from the analyses of SPS energy data on strange baryon and antibaryon
production \cite{BM_SPS,heinz}
and of AGS energy reactions \cite{BM_AGS}.
The time scale turns out to be about $10\sim 100\> \rm fm/c$ at $T=180\>\rm
MeV$ except for the AGS experiments. 
This is too large to change the number of pions in expanding hot 
hadronic matter. For AGS experiments with $\mu_B$= 516 MeV we find  that
$\tau_{ch}\sim 5 \>\rm fm/c$ at $T\sim 150$ MeV. This implies that the baryon
effect is important for AGS experiments but negligible in case of SPS
experiments. In the following we, therefore, ignore the corrections due to 
the presence of baryons.

\section{Chemical equilibrium with finite pion chemical potential}

So far we have considered the characteristic time scale 
for chemical equilibration of pions when the system is slightly out of 
chemical equilibrium.  
This is the case either when the hadronic system evolves from a fully
equilibrated quark-gluon plasma phase and hadronizes without any changes in
chemical properties or when the hadronic system directly is produced in
thermodynamic equilibrium state. 
However, if the system is out of chemical equilibrium initially,
the reaction rate for the pion number changing processes 
will be affected by the finite pion chemical potential. 

We, therefore,  
extend the previous definition for the chemical relaxation time to the case
with finite chemical potential and define
\begin{equation}
{1\over \tau_{ch}} \equiv {-\delta n\over [n(T,\mu_\pi)-n^0(T)]}
                        {dn(T,\mu_\pi)\over dt},
\label{eq.3.1}
\end{equation}
which is the ratio of the pion number changing rate over the  the excess 
(or lack) of pions as compared to equilibrium.  
In the limit of vanishing chemical potential, $\mu_\pi \to 0$, the resulting
chemical relaxation time coincides with that derived in the previous section.

\subsection{Chemical relaxation time of pions: out-of equilibrium}

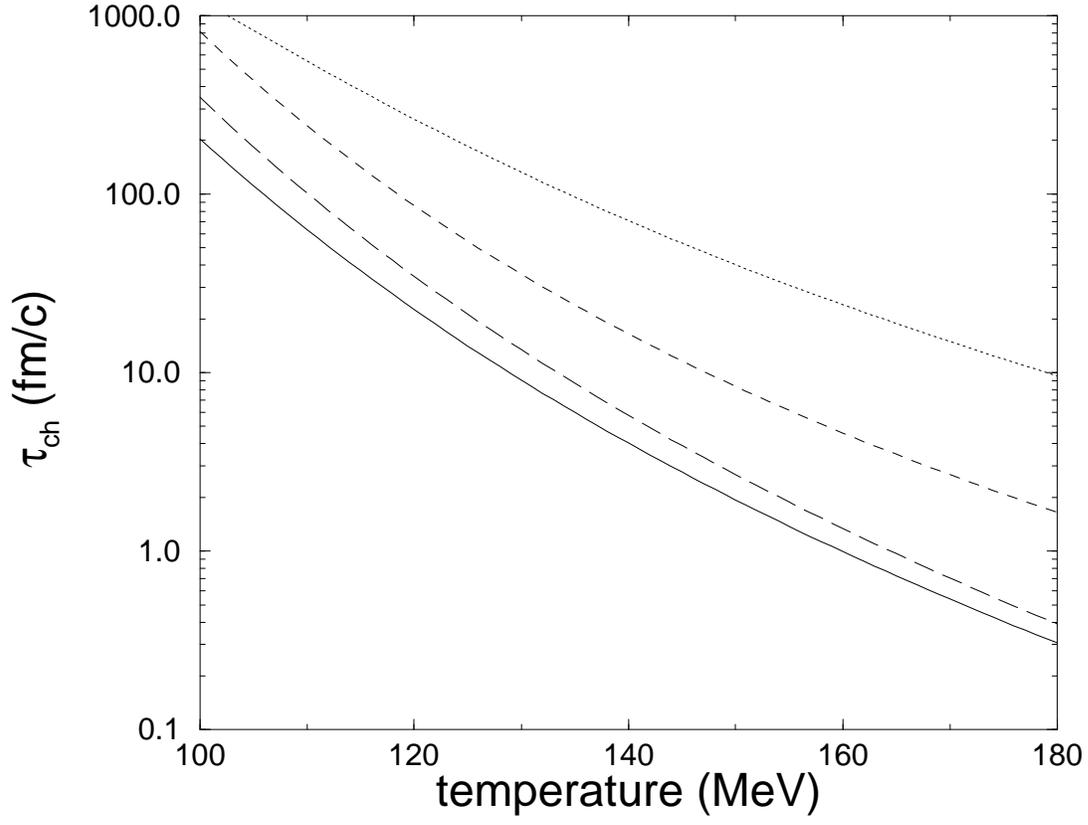
\begin{figure}
\caption{Chemical relaxation time of pions
in hot hadronic matter with finite pion chemical potential. 
The contributions from the reaction $\rho\rho\leftrightarrow\pi\pi$ 
              (long dashed curve),
              $\pi a_1\leftrightarrow\pi\pi$ (dashed curve) and 
              $\pi\pi\pi\pi\leftrightarrow\pi\pi$ (dotted curve) are shown.
              The solid
              curve is for the total relaxation time.
The results are obtained with $\mu_\pi=100 \>\rm MeV$.}
\label{pion_number_1}
\end{figure}

In the previous section  we found that for a pure pion gas
the chemical equilibration rate evaluated near chemical equilibrium 
is too long to to be effective.
If the system is far away from the chemical equilibrium, 
then the annihilation (production) reaction rates depend on the value of the
chemical potential. For the reaction $\pi\pi\pi\pi\to\pi\pi$
with finite chemical potential of pions, the chemical relaxation time 
Eq.~(\ref{eq.3.1}) is given by 
\begin{equation}             
{1\over \tau_{ch}^\pi}=-{2\over n^0_\pi(T)} 
                    e^{2\mu_\pi/T}(1-e^{2\mu_\pi/T})
                    I_0(\pi\pi\pi\pi\rightleftharpoons\pi\pi;T)
                   /(e^{\mu_\pi/T}-1),
\end{equation}
where we take Boltzmann approximation for the distribution of pions.
The calculated chemical relaxation time is shown as dotted 
line in Fig.~\ref{pion_number_1} 
for $\mu_\pi=100\, \rm MeV$. When the system is  out of equilibrium the 
number changing process becomes considerably faster -- 
about a factor of 5 at $T = 150 \, \rm MeV$ -- than in the previous case  
where the system was near equilibrium (note the different scale on the y-axis 
of Fig.~\ref{pion_number_1}). However, the reaction rate     
is still large compared to the size of the system, 
$\tau^\pi_{ch}\sim 10$ fm/c  at $T=180\,\rm MeV$.
Thus, in a hadronic system consisting of only pions, 
the number of pions ceases to change at the early stage of the expansion 
even if the system is out of equilibrium at the beginning of the evolution.

Once we include resonances additional processes help to change the pion
number. Again we assume that the decay processes such as
$\rho\rightleftharpoons\pi\pi$ and $a_1\rightleftharpoons\pi\rho$ 
are fast enough to 
maintain the {\it relative} chemical equilibrium even if pions have a 
finite chemical potential.
From Eq.~(\ref{par_number}) we can see that the pion number is not changed 
by these reactions. The pion number will be
changed, however, by the reactions such as $\rho\rho\to\pi\pi$ 
and $\pi a_1\to\pi\pi$;
these reactions actually mean that four pions decay into two pions. 
The chemical relaxation time for the corresponding reactions 
is given by
\begin{equation}
{1\over \tau^\pi_{ch}}=-{2\over n_\pi^0(T)}
(e^{2 \mu_\pi/T}-e^{2 \mu_\rho/T})
I_0(\rho\rho\rightleftharpoons\pi\pi;T)/(e^{\mu_\pi/T}-1),
\end{equation}
with $\mu_\rho=2\mu_\pi$. 
We also get the similar relation for the $\pi a_1\to\pi\pi$;
\begin{equation}
{1\over \tau^\pi_{ch}} = -{2\over n_\pi^0(T)}
(e^{2 \mu_\pi/T}-e^{(\mu_\pi+\mu_{a_1})/T})
I_0(\pi a_1\rightleftharpoons\pi\pi;T)/(e^{\mu_\pi/T}-1),
\end{equation}
where $\mu_{a_1}=3\mu_\pi$.

We show the relaxation time with 
$\mu_\pi=100\rm\, MeV$ in Fig.~\ref{pion_number_1}.   
The total chemical relaxation time for $\pi\pi\pi\pi\rightleftharpoons\pi\pi$ 
is given by the sum for all possible channels. 
With a finite chemical potential of pions, the chemical
relaxation time of pions now becomes 
short compared to the size of the system. Thus the number 
of pions would be changed in hot hadronic matter even near the thermal 
freeze-out temperature.
The oversaturated pion number will be
reduced mainly by the reactions $\rho\rho\to\pi\pi$ and $\pi a_1\to\pi\pi$.
Especially axial vector mesons very easily interact
with pions and annihilate into two pions.  
However, we should note that the thermal 
freeze-out temperature
is also reduced when we include a finite pion chemical potential
\cite{thermal}.

With non-zero pion
chemical potential one also expects an  oversaturation of vector and axial 
vector mesons with the relation $\mu_\rho=2\mu_\pi$ and $\mu_{a_1}=3\mu_\pi$, 
respectively. The above result, of course,  implies that also 
the number of vector and axial vector mesons changes as quickly as that of
the pions.

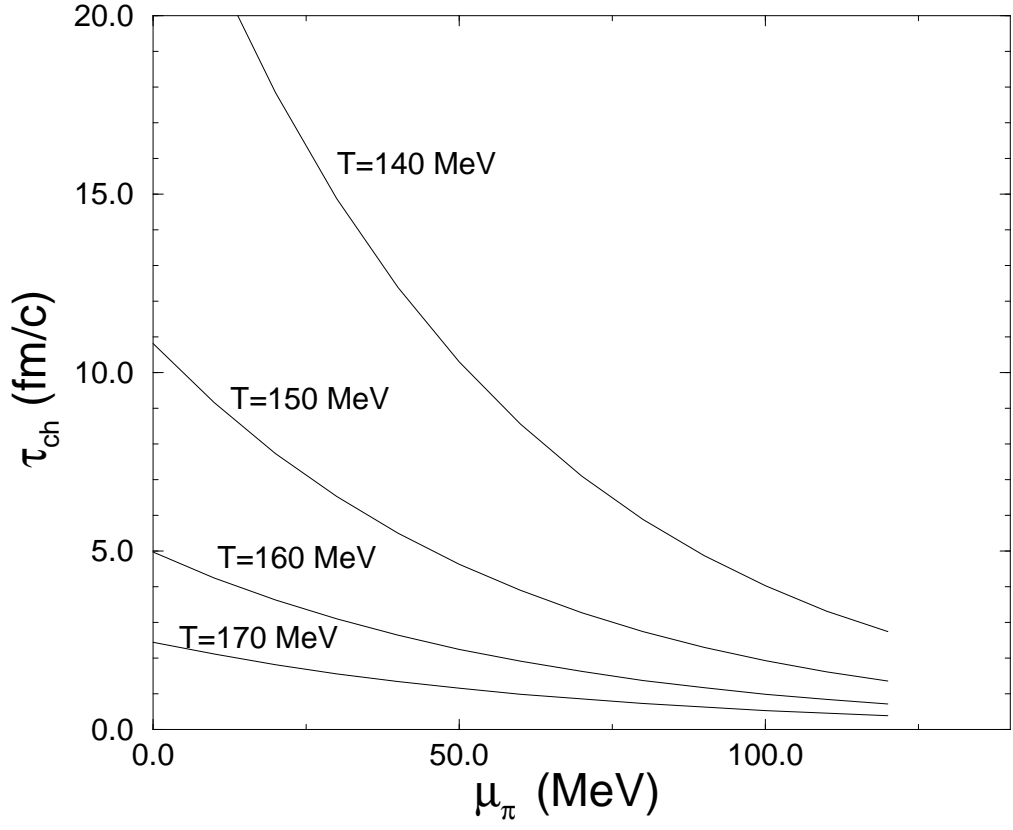
\begin{figure}
\caption{Chemical relaxation time of pions in hot hadronic matter as function
of pion chemical potential. 
We consider four different temperatures; from top 
$T = 140,\> 150,\> 160, \>170 $ MeV.}
\label{t_ch}
\end{figure}

In Fig.~\ref{t_ch} we show how the total 
chemical relaxation time is changed with
pion chemical potential at fixed temperature. Even at $T=150$ MeV, the pion
relaxation time is comparable to the size of the hot system as 
long as the pions
have a finite chemical potential; $\tau_{ch} \approx 5\>$ fm/c at $\mu_\pi= 50$
MeV. 
If the hadronic system is produced out of equilibrium, for example with 
$\mu_\pi=100\,\rm MeV$, the excess of pions will be reduced by the 
inelastic reactions involving vector and axial vector mesons.  
These number changing processes will lead to the decrease of pion chemical 
potential and finally cease to be effective as the chemical potential is 
reduced below a certain value. Therefore, given a freeze-out temperature of
about $150 \, \rm MeV$ we would expect the pion chemical potential not to
exceed a value of $\simeq 50 \, \rm MeV$. 

\subsection{Chemical relaxation time of pions: in equilibrium}
In the previous section we have demonstrated that the chemical relaxation time
depends on both the temperature as well as the chemical potential. In order
to give a realistic estimate of the chemical relaxation time, one thus needs to
know the chemical potential as a function of temperature. The correct
treatment, of course, requires the complete solution of all kinetic equations 
including the expansion of the system. 
This is best done in a transport approach
and will be addressed in a separate work. In order to give a rough estimate let
us separate the problem into two pieces. First of all, we determine the
chemical potential as a function of temperature, assuming the {\em absence} of
any number changing processes. Given that, we can then evaluate the chemical
relaxation time using our previous results.

In order to determine the chemical potential as a function of temperature, we
assume that the system initially (after hadronization) is in chemical
equilibrium. We furthermore ignore all true number changing processes.
In this case we have only two possible reactions in hadronic phase;
elastic pion scattering which
maintains the thermal equilibrium and the decay of the excited states,
e.g. $\rho\to\pi\pi$ and $a_1\to\pi\rho$. 
The local configuration of pions is then described 
by temperature and a single chemical potential, $\mu$. 

Since the system still remains in thermal
equilibrium, the state can be characterized by a maximum of the entropy
consistent with the conservation laws of energy, momentum and of relevant
particle numbers. Since we neglect dissipative, i.e. pion number
changing, processes this implies that entropy
as well as the pion number is conserved in the expansion. 
Thus one can assume that the ratio of the effective pion number density $\bar
n_\pi=n_\pi+2n_\rho+3n_{a_1}+\cdots$  to entropy density 
$s$ remains constant during  the expansion after
hadronization until  freeze-out,
\begin{equation}
{s\over\bar n_\pi}=constant.
\end{equation}
As a result of keeping ${s/\bar n_\pi}$  constant, the chemical potential will 
rise. This is due to the overpopulation of pionic states due to the decay of
the resonances. It has been estimated that pion chemical potential is about 
86 MeV at freeze-out temperature \cite{bebie}.
To be specific, we assume that the particles are distributed according to
\begin{equation}
f_i={g_i\over\exp(E_i-\mu_i)/T-1},
\end{equation}
where $i=\pi,\,\rho,\, a_1 \cdots$ and $g_i$ is the spin and isospin
degeneracy factor and $E_i=\sqrt{p^2+m_i^2}$.
The entropy density is given by 
\begin{equation}
s={\partial P\over\partial T}\Biggl\vert_{\mu=const},
\end{equation}
where the pressure is 
\begin{eqnarray}
P=T\sum_ig_i\int{d^3p\over(2\pi)^3}
  \ln\left\{1+\exp\left({\mu_i-E_i\over T}\right)\right\}.
\end{eqnarray}

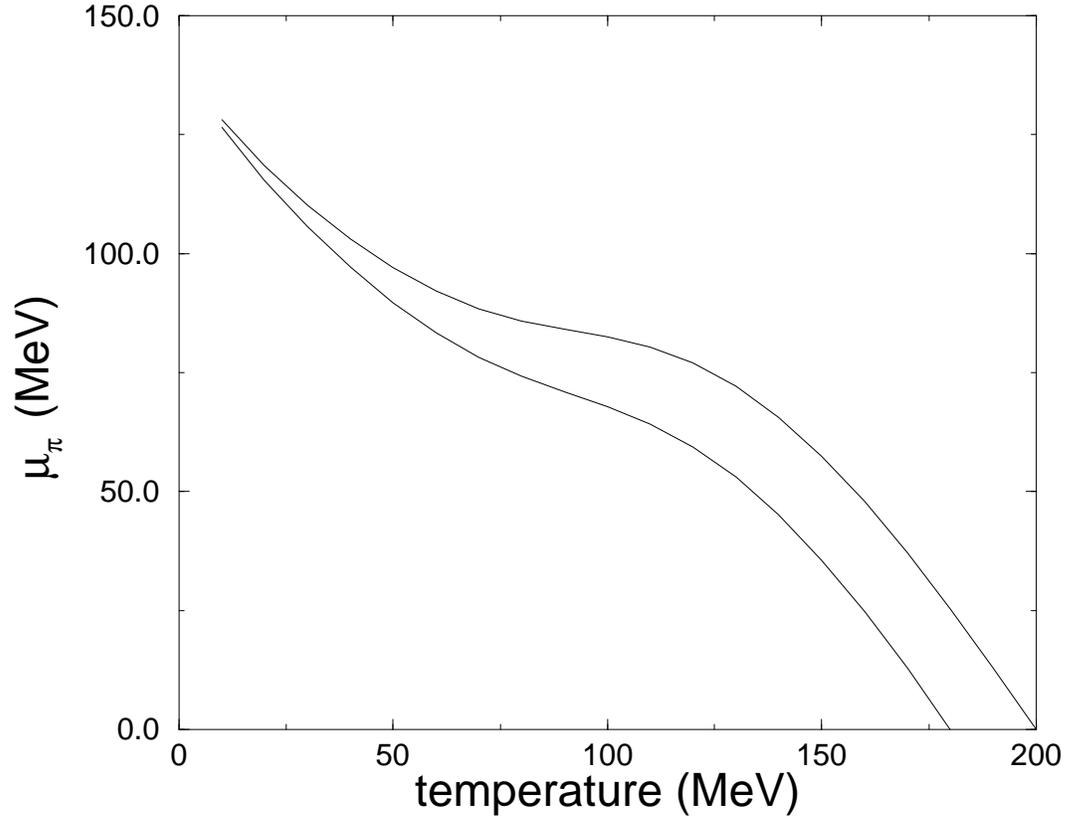
\begin{figure}
\caption{Temperature dependence of the pion chemical potential: We assume two
different  initial (hadronization) temperatures,  $T_h=$ 180 (lower curve) and 
200 MeV (upper curve).}
\label{chem_poten}
\end{figure}

Assuming that the ratio of
entropy density of the system to effective pion number density is constant
throughout expansion, we can obtain the temperature dependence of the
pion chemical potential as shown in Fig.~\ref{chem_poten} \cite{bebie}.
Here  we take two different initial (hadronization) temperatures, 
$T_h=$180 and 200 MeV,
where the chemical potential is assumed to be zero, $\mu_\pi=0$. 

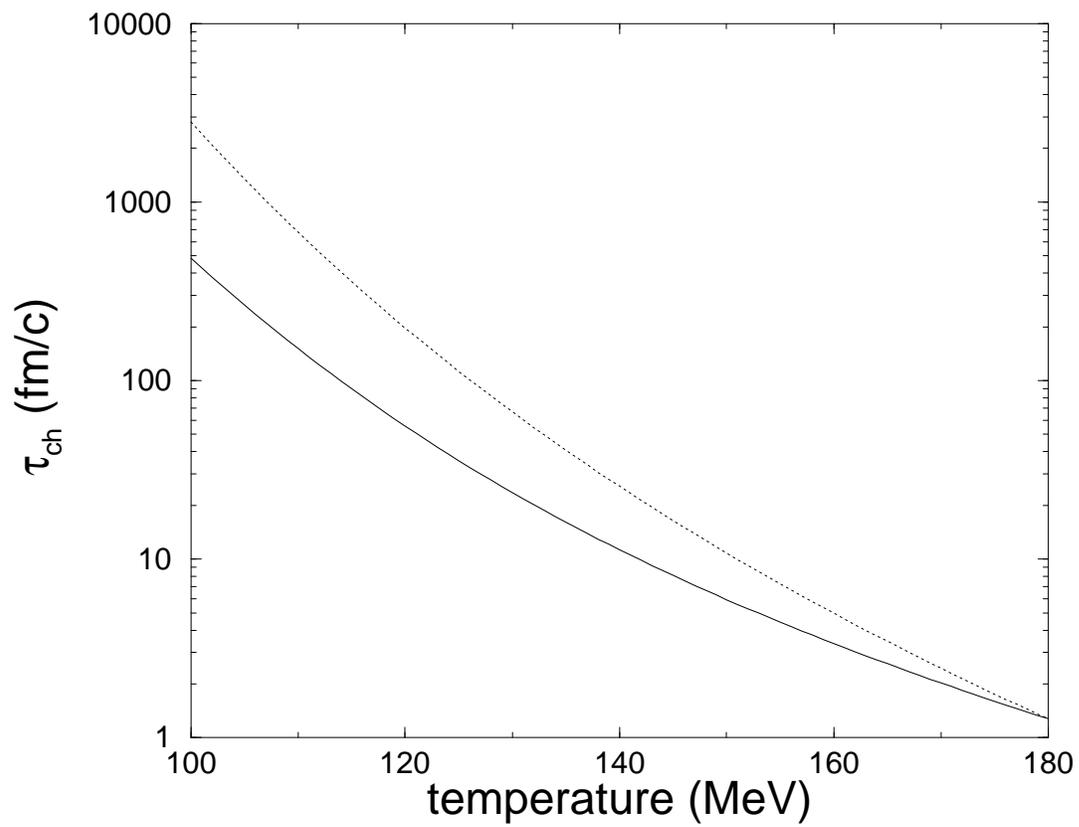
\begin{figure}
\caption{Chemical relaxation time of pions 
with pion chemical potential shown in Fig.~3. Here we assume that 
$T_h$=180 MeV.}
\label{pion_number_2}
\end{figure}

With the temperature dependence of the pion chemical potential 
we can calculate the total chemical relaxation time based on the reaction 
 $\pi\pi\pi\pi\to\pi\pi$, $\rho\rho\to\pi\pi$ and $a_1\pi\to\pi\pi$. 
We show the result 
in Fig.~\ref{pion_number_2}. 
At the initial temperature we, of course,
have the same value for the chemical relaxation time as that 
obtained in the limit $\mu_\pi\to 0$.
As the system expands and pions develop the chemical potential
the relaxation time becomes shorter than that obtained
near equilibrium (as shown by dotted line). 
We find that the $\tau_{ch}$ is reduced by half at $T= 150$ MeV 
($\tau_{ch}\sim 5$ fm/c) 
due to the pion chemical potential. However, even with the induced pion 
chemical potential the chemical relaxation time is considerably larger than
the thermal relaxation time at the same temperature.

\section {Summary}

We have studied the thermal and chemical relaxation time scales of pions in hot
hadronic matter with an effective chiral Lagrangian. 
From the explicit calculation we show
that pions in hot hadronic matter are in a phase where 
elastic collisions rates are very fast compared to typical 
expansion rates of the system.
For chemical equilibration the dominant contribution comes from 
the inelastic collision involving $a_1$ mesons. 
Comparing with previous
calculations \cite{goity} based on chiral perturbation theory, the inclusion of
the resonances has reduced the chemical relaxation time by about a factor of
10. When we neglect the formation time of these resonances, the resulting 
chemical relaxation time of pions is 10 fm/c at $T=150$ MeV.  This value is 
comparable to the size of the hot system produced the collision
of large nuclei.  

Given a system size of $5\sim 10 \, \rm fm$ we obtain a thermal freeze-out 
temperature which is small compared to those extracted from experiments
\cite{BM,heinz}. This might be due to  
flow effects which lead to smaller effective system sizes.
If we take the thermal freeze-out temperature be about 150 MeV, then the 
freeze-out size of the system would be $2\sim 3$ fm. On the other hand 
the chemical relaxation time for a system of this size would be 
 $T=170\, \rm MeV$. This implies 
that chemical freeze-out of pions happens at considerably higher temperatures
than thermal freeze-out. This result is somewhat at variance with the findings
of \cite{BM_SPS,rafelski}, where pion spectra and particle abundances 
could be reproduced assuming the same freeze-out conditions.
In order to properly assess the magnitude and importance of this
discrepancy, a detailed transport calculation including all the 
number changing processes presented here is needed.   

We also have studied 
the effect of baryons on the chemical relaxation time of pions. 
Since the effect of baryons is suppressed by their large mass, we consider only
low laying baryons, $N(938)$ and $\Delta(1236)$. To estimate the relaxation
time
we use the phenomenological cross section for $NN\rightleftharpoons N\Delta$. 
The effect of baryons is very small and can be neglected in SPS experiments.
However, it becomes important in AGS experiments where the baryon
chemical potential is much larger than that in the SPS experiments.  

We have extend the definition for the chemical relaxation time to a
system of pions  out of equilibrium with $\mu_\pi\ne 0$. 
In this case it is
possible that the overpopulated pions will be reduced by the inelastic
reactions involving vector and axial vector mesons. 
At $T=150$ with $\mu_\pi=100\,\rm MeV$ 
the relaxation time is about 2 fm/c which is certainly 
comparable to the system size 
at freeze-out. However, we should note that the thermal freeze-out
temperature is also reduced when we include a finite pion chemical potential
\cite{thermal}. 

In order to make contact with reality, we have determined the pion chemical
potential as a function of temperature assuming  isentropic expansion while 
ignoring  number changing processes.
Based on this relation, we could give a somewhat more realistic estimate of the
actual chemical relaxation time. We found that  in this case the chemical
relaxation time is about $5 \, \rm fm/c$ at $T=$150 MeV. For systems which are
larger than $5 \, \rm fm$ we, therefore, expect that the chemical potential
at freeze-out will be considerably smaller than the value of $86 \, \rm MeV$,
which has been obtained without taking number changing processes into account.

In conclusion, as long as the effective system size is not considerably smaller
than $5 \, \rm fm$, a buildup of a pion chemical potential larger than $100
\, \rm MeV$ would be very difficult to understand. 
At the same time, we also predict a
considerable difference between the chemical and thermal freeze-out
temperatures. To which extent that is reflected in the data needs to be
investigated within a transport calculation.

Future work will concentrate on a transport theoretical calculation of the
chemical equilibration.  We also plan to extent the present  study to include 
the reactions involving  strange particles.

\acknowledgements
This work supported by the Director, Office of Energy Research, Office of High
Energy and Nuclear Physics, Division of Nuclear Physics, Division of Nuclear
Sciences, of the U. S. Department of Energy under Contract 
No. DE-AC03-76SF00098.
\eject

\newpage

\section{Appendix A: Effective Chiral Lagrangian}

We consider an effective chiral Lagrangian with vector and axial-vector meson
fields which are introduced as massive Yang-Mills fields \cite{mym};
\begin{eqnarray}
{\cal L} &=&  {1\over4}f_\pi^2{\rm Tr}[D_\mu U D^\mu U^\dagger]
             +{1\over4}f_\pi^2{\rm Tr}M(U+U^\dagger-2)\cr
         & & -{1\over2}{\rm Tr}[ F^L_{\mu\nu}F^{L\mu\nu}
                                +F^R_{\mu\nu}F^{R\mu\nu}]             
                 +m_0^2{\rm Tr}[A^L_\mu A^{L\mu}+A^R_\mu A^{R\mu}]\cr
         & & -i\xi {\rm Tr}[D_\mu U D_\nu U^\dagger F^{L\mu\nu} 
                          +D_\mu U^\dagger D_\nu U F^{R\mu\nu}]\cr
         & & +\sigma {\rm Tr} [F^L_{\mu\nu}UF^{R\mu\nu}U^\dagger],
\end{eqnarray}
where $U$ is related to the pseudoscalar fields $\phi$ by
\begin{equation}
U=\exp\left[{i\sqrt{2}\over f_\pi}\phi\right],
\quad\phi=\sum_{a=1}^3\phi_a{\tau_a\over\sqrt{2}},
\end{equation}
and $A^L_\mu (A^R_\mu)$ are left(right)-handed vector fields.
The covariant derivative acting on $U$ is given by 
\begin{equation}
D_\mu U=\partial_\mu U-igA^L_\mu U-igUA^R_\mu,
\end{equation}
and $F^L_{\mu\nu}\,(F^R_{\mu\nu})$ is the field tensor 
of left(right)-handed vector fields. The $A^L_\mu$ and $A^R_\mu$
can be written in terms of vector ($V_\mu$) and 
axial-vector fields ($A_\mu$) as 
\begin{equation}
A_\mu^L={1\over2}(V_\mu-A_\mu), \quad A_\mu^R={1\over2}(V_\mu+A_\mu).
\end{equation}
With parameters $g=10.3063,\>\sigma=0.3405,\>\xi=0.4473$ \cite{song} we can
well describe the properties of vector and axial vector mesons.

From the Lagrangian we have 
\begin{eqnarray}
{\cal L}^{(3)}_{V\phi\phi}&=
&{ig\over 2} {\rm Tr}{\partial_\mu \phi[V^\mu,\phi]}
+{ig\delta\over 2m_V^2} {\rm Tr}{(\partial_\mu V_\nu
                    -\partial_\nu V_\mu) \partial^\mu\phi \partial^\nu\phi},\cr
{\cal L}^{(3)}_{VVV}&=
 & {ig\over 2}{\rm Tr}{(\partial_\mu V_\nu-\partial_\nu V_\mu) V^\mu V^\nu},\cr
{\cal L}^{(3)}_{VA\phi}&=&{ig\over 2}{1\over f_\pi}\biggl\{
 \eta_1{\rm Tr}(\partial_\mu V_\nu-\partial_\nu V_\mu)
                [A^\mu,\partial^\nu\phi]
+\eta_2{\rm Tr}(\partial_\mu A_\nu-\partial_\nu A_\mu)
                [\partial^\mu V^\nu, \phi]\biggr\},
\end{eqnarray}
where $\delta,\>\eta_1$, and $\eta_2$ are given by the parameters in the 
Lagrangian and have the values $0.347,\, 0.279,\, 0.060$, respectively.

For the four point vertex we have 
\begin{eqnarray}
{\cal L}^{(4)}_{VV\phi\phi}&=& -{1\over8}g^2{\rm Tr}[V_\mu,\phi]^2
\nonumber\\
&&
+{c_1\over f_\pi^2}{\rm Tr}[(\partial_\mu V_\nu-\partial_\nu V_\mu)\phi
             (\partial^\mu V^\nu-\partial^\nu V^\mu)\phi
            -(\partial_\mu V_\nu-\partial_\nu V_\mu)^2\phi^2]
\nonumber\\
&&
+{c_2\over f_\pi^2}{\rm Tr}[(\partial_\mu V_\nu-\partial_\nu V_\mu)
             (\partial^\mu\phi [V^\nu,\phi]
             +[V^\mu,\phi]\partial^\nu\phi)]
\nonumber\\
&&
+{c_2\over f_\pi^2}{\rm Tr}[V_\mu,V_\nu]\partial^\mu\phi\partial^\nu\phi,
\end{eqnarray}
and also have 
\begin{eqnarray}
{\cal L}^{(4)}_{A\phi^3}&=&
 {d_1\over f_\pi}{\rm Tr}A_\mu(\phi^2\partial^\mu\phi+\partial^\mu\phi\phi^2)
\nonumber\\
&&
+{d_2\over f_\pi}{\rm Tr}A_\mu\phi\partial^\mu\phi\phi
\nonumber\\
&&
+{d_3\over f_\pi^3}
{\rm Tr}[A_\mu,\partial_\nu\phi][\partial^\mu\phi,\partial^\nu\phi]
\nonumber\\
&&
+{d_4\over f_\pi^3}{\rm Tr}(\partial_\mu A_\nu-\partial_\nu A_\mu)
            (\partial^\mu\phi\partial^\nu\phi\phi -
             \phi\partial^\mu\phi\partial^\nu\phi),
\end{eqnarray}
where $c$'s and $d$'s are given by the parameters in the Lagrangian. 
With parameters given in Ref.~\cite{song} 
we have the values $0.00497,\, -0.00408$ for 
$c_1$ and $c_2$, respectively and
$d_1=-2.08,\,d_2=-2d_1,\,d_3=0.078,\,{\rm and}\, d_4=-0.0147$.

We also include the gauged Wess-Zumino term 
in the effective Lagrangian to describe the non-Abelian anomaly structure
of QCD \cite{witten}, which
leads to an anomalous interaction among a pseudoscalar meson and two
vector mesons, 
\begin{equation}
{\cal L}_{VVP}=-{3g^2\over16\sqrt{2}\pi^2f_\pi}\epsilon^{\mu\nu\alpha\beta}
                   {\rm Tr}(\partial_\mu V_\nu\partial_\alpha V_\beta P),
\label{vvp}
\end{equation}
where $\epsilon^{\mu\nu\alpha\beta}$ is the antisymmetric Levi-Civita tensor 
with $\epsilon^{0123}=1$.
From Eq.~(\ref{vvp}) we have 
\begin{equation}
{\cal L}_{\omega\pi\rho}=-g_\omega\epsilon^{\mu\nu\alpha\beta}
                         \partial_\mu \omega_\nu
                         \partial_\alpha\rho_\beta\cdot\pi,
\end{equation}
with 
\begin{equation}
g_\omega=\left({3g^2\over 16\pi^2f_\pi}\right).
\end{equation}

\section {Appendix B: Inelastic cross section}

In this appendix we summarize the invariant matrix element 
for each inelastic collision considered in section (II.C). The interaction for 
each process is obtained from the effective Lagrangian.

\subsection* {B.1  $\quad\pi+\pi\to\rho+\rho$}

\subsubsection*{B.1.1.  $\quad\pi^{+}+\pi^{-}\to\rho^{+}+\rho^{-}$;
                        $\pi^{+}+\pi^{0}\to\rho^{+}+\rho^{0}$}

The related diagrams are shown in Fig.~7. The explicit expression for
the invariant matrix element for each reaction is given by 
\begin{eqnarray}
{\cal M}^{(a)} &=& i\left[{g\over\sqrt{2}}
                    \Biggl(1-{\delta\over 2}\Biggr)\right]^2
                    {4p_{1\mu}p_{2\nu}\over (p_1-p_3)^2-m_\pi^2}
                    \epsilon^{\mu}_\rho(p_3)\epsilon^{\nu}_\rho(p_4),
\\[12pt]
{\cal M}^{(b)} &=& i\left({g\over\sqrt{2}}\right)^2
                   {1\over (p_1+p_2)^2-m_\rho^2}
                   \left(1-{\delta \over 2m_\rho^2}(p_1+p_2)^2\right)
\nonumber\\
&&\times \biggl[(p_1-p_2)\cdot(p_3-p_4)g_{\mu\nu}
                +2p_{4\mu}(p_{1\nu}-p_{2\nu})
                -2(p_{1\mu}-p_{2\mu})p_{3\nu}\biggr]
         \epsilon^{\mu}_\rho(p_3)\epsilon^{\nu}_\rho(p_4),
\\[12pt]
{\cal M}^{(c)} &=& i\left({g\over\sqrt{2}}\right)^2
                   \Biggl\{ [1+4{c_1\over f_\pi^2} p_3\cdot p_4
-2{c_2\over f_\pi^2}(p_1\cdot p_4+p_2\cdot p_3)]g_{\mu\nu}
\nonumber\\
&&\qquad\qquad
          -4{c_1\over f_\pi^2}p_{4\mu}p_{3\nu}
          +2{c_2\over f_\pi^2}( p_{4\mu}p_{1\nu}+p_{2\mu}p_{3\nu}
                -p_{1\mu}p_{2\nu}+p_{2\mu}p_{1\nu})\Biggr\}
          \epsilon^{\mu}_\rho(p_3)\epsilon^{\nu}_\rho(p_4),
\end{eqnarray}
where $c_1$ and $c_2$ are given in previous section.

The invariant scattering matrix element is then given by the sum as
\begin{eqnarray}
{\cal M}(\pi^{+}+\pi^{-}\to\rho^{+}+\rho^{-})
&=&{\cal M}(\pi^{+}+\pi^{0}\to\rho^{+}+\rho^{0})\cr
&=&{\cal M}^{(a)}+{\cal M}^{(b)}+{\cal M}^{(c)}.
\end{eqnarray}

\subsubsection*{B.1.2. $\quad\pi^{+}+\pi^{-}\to\rho^{0}+\rho^{0}$;
                       $\pi^{0}+\pi^{0}\to\rho^{+}+\rho^{-}$}

The related diagrams are shown in Figs.~8. The explicit expression for
the invariant matrix element for each reaction is given by 
\begin{eqnarray}
{\cal M}^{(a)} &=& {-i\over2}\left[{g\over\sqrt{2}}
                   \Biggl(1-{\delta\over 2}\Biggr)\right]^2     
                   {4p_{1\mu}p_{2\nu}\over (p_1-p_3)^2-m_\pi^2}
                   \epsilon^{\mu}_\rho(p_3)\epsilon^{\nu}_\rho(p_4),
\\[12pt]
{\cal M}^{(b)} &=& {-i\over2}\left[{g\over\sqrt{2}}
                   \Biggl(1-{\delta\over 2}\Biggr)\right]^2
                   {4p_{2\mu}p_{1\nu}\over (p_3-p_2)^2-m_\pi^2}
                    \epsilon^{\mu}_\rho(p_3)\epsilon^{\nu}_\rho(p_4),
\\[12pt]
{\cal M}^{(c)} &=& i\left({g\over\sqrt{2}}\right)^2
                    \Biggl\{ [1+4{c_1\over f_\pi^2}p_3\cdot p_4
-{c_2\over f_\pi^2}(p_1+p_2)^2]g_{\mu\nu}
\nonumber\\
&&\qquad\qquad\qquad\qquad
          -4{c_1\over f_\pi^2} p_{4\mu}p_{3\nu}
          +2{c_2\over f_\pi^2} p_{4\mu}p_{3\nu}\Biggr\}
 \epsilon^{\mu}_\rho(p_3)\epsilon^{\nu}_\rho(p_4).
\end{eqnarray}
The invariant scattering matrix element is given by 
\begin{eqnarray}
{\cal M}(\pi^{+}+\pi^{-}\to\rho^{0}+\rho^{0})
&=&{\cal M}(\pi^{0}+\pi^{0}\to\rho^{+}+\rho^{-})\cr
&=&{\cal M}^{(a)}+{\cal M}^{(b)}+{\cal M}^{(c)}.
\end{eqnarray}

\subsection* {B.2 $\quad\pi+\pi\to\pi+\omega$}
The related diagrams are shown in Fig.~9. The explicit expression for
the invariant matrix element is 
\begin{equation}
{\cal M}(\pi+\pi\to\pi+\omega)
={\cal M}^{(a)}+{\cal M}^{(b)}+{\cal M}^{(c)},
\end{equation}
where
\begin{eqnarray}
{\cal M}^{(a)} &=& ig_{\omega\rho\pi}{g\over\sqrt{2}}
                   \left(1-{\delta\over2m_\rho^2}(p_1-p_3)^2\right)
                   \epsilon^{\mu\nu\alpha\beta}
                   {(p_{1\alpha}+p_{3\alpha})(p_{1\beta}-p_{3\beta})
                    \over(p_1-p_3)^2-m_\rho^2}
                   \epsilon_{\mu}^\omega(p_4)p_{4\nu},
\\[12pt]
{\cal M}^{(b)} &=& ig_{\omega\rho\pi}{g\over\sqrt{2}}
                   \left(1-{\delta\over2m_\rho^2}(p_3-p_2)^2\right)
                   \epsilon^{\mu\nu\alpha\beta}
                   {(p_{3\alpha}+p_{2\alpha})(p_{1\beta}-p_{4\beta})
                   \over(p_1-p_4)^2-m_\rho^2}
                   \epsilon_{\mu}^\omega(p_4)p_{4\nu},
\\[12pt]
{\cal M}^{(c)} &=& -ig_{\omega\rho\pi}{g\over\sqrt{2}}
                   \left(1-{\delta\over2m_\rho^2}(p_1+p_2)^2\right)
                   \epsilon^{\mu\nu\alpha\beta}
                   {(p_{1\alpha}-p_{2\alpha})(p_{1\beta}+p_{2\beta})
                    \over(p_1+p_2)^2-m_\rho^2}
                   \epsilon_{\mu}^\omega(p_4)p_{4\nu}.
\end{eqnarray}

\subsection* {B.3 $\quad\pi+a_1\to\pi+\pi$}

\subsubsection* {B.3.1 $\quad\pi^{+}+a_1^{0}\to\pi^{0}+\pi^{+}$;
                       $\pi^{+}+a_1^{-}\to\pi^{+}+\pi^{-}$}
The related diagrams are shown in Fig.~10.
The explicit expression for
the invariant matrix element for each reaction is given by 
\begin{eqnarray}
{\cal M}^{(a)}&=&-\left({g\over\sqrt{2}}\right)^2{1\over f_\pi}
                      {1\over (p_1-p_3)^2-m_\rho^2}
                \left(1-{\delta\over 2m_\rho^2} (p_1-p_3)^2\right)
\nonumber\\
&&\times\Biggl[\biggl(\eta_1 p_4\cdot(p_1-p_3)-\eta_2p_2\cdot(p_1-p_3)\biggr)
             (p_{1\,\mu}+p_{3\,\mu})
\nonumber\\
&&\qquad\qquad
      -\biggl(\eta_1 p_4\cdot(p_1+p_3)-\eta_2p_2\cdot(p_1+p_3)\biggr)
             (p_{1\,\mu}-p_{3\,\mu})\Biggr]
       \epsilon^{\mu}_{a_1}(p_2),
\\[12pt]
{\cal M}^{(b)}&=&-\left({g\over\sqrt{2}}\right)^2{1\over f_\pi}
                      {1\over (p_1+p_2)^2-m_\rho^2}
                \left(1-{\delta\over 2m_\rho^2} (p_1+p_2)^2\right)
\nonumber\\
&&\times\Biggl[\biggl(\eta_1 p_1\cdot(p_1+p_2)+\eta_2p_2\cdot(p_1+p_2)\biggr)
             (p_{3\,\mu}-p_{4\,\mu})
\nonumber\\
&&\qquad\qquad
      -\biggl(\eta_1 p_1\cdot(p_3-p_4)+\eta_2p_2\cdot(p_3-p_4)\biggr)
             (p_{3\,\mu}+p_{4\,\mu})\Biggr] 
       \epsilon^{\mu}_{a_1}(p_2),
\\[12pt]
{\cal M}^{(c)}_1&=& {1\over f_\pi^3}\Biggl[
 \biggl(2d_2 f_\pi^2 -2d_3 p_3\cdot p_4+2d_4 p_2\cdot p_3\biggr) p^\mu_1
-\biggl(d_2 f_\pi^2 +2d_3 p_1\cdot p_3+2d_4 p_2\cdot p_3\biggr) p^\mu_4
\nonumber\\
&&\qquad\qquad
-\biggl(2 d_1 f_\pi^2-d_2 f_\pi^2 -2d_3 p_1\cdot p_4-2d_4 (p_4-p_1)\cdot
p_2\biggr) p^\mu_3 
\Biggr]\epsilon^{\mu}_{a_1}(p_2),
\\[12pt]
{\cal M}^{(c)}_2&=&{1\over f_\pi^3} \Biggl[
 \biggl(2d_1 f_\pi^2 +2d_3 p_3\cdot p_4-2d_4 p_2\cdot p_3\biggr) p^\mu_1
-\biggl(2d_1 f_\pi^2 -2d_3 p_1\cdot p_3-2d_4 p_2\cdot p_3\biggr) p^\mu_4
\nonumber\\
&&\qquad\qquad-\biggl(2d_2 f_\pi^2 +4d_3 p_1\cdot p_4+2d_4 (p_4-p_1)\cdot
p_2\biggr) p^\mu_3 
\Biggr]\epsilon^{\mu}_{a_1}(p_2),
\end{eqnarray}
where $d_i$'s are given by parameters in the Lagrangian. See Appendix A.

The invariant matrix element is given by 
\begin{eqnarray}
{\cal M}(\pi^{+}+a_1^{0}\to\pi^{0}+\pi^{+})
&=&{\cal M}^{(a)}+{\cal M}^{(b)}+{\cal M}^{(c)}_1,
\nonumber\\
{\cal M}(\pi^{+}+a_1^{-}\to\pi^{+}+\pi^{-})
&=&{\cal M}^{(a)}+{\cal M}^{(b)}+{\cal M}^{(c)}_2.
\end{eqnarray}

\subsubsection* {B.3.2 $\quad\pi^{+}+a_1^{-}\to\pi^{0}+\pi^{0}$;
                       $\pi^{+}+a_1^{+}\to\pi^{+}+\pi^{+}$}
The related diagrams are shown in Fig.~11.
The explicit expression for
the invariant matrix element for each reaction is given by 
\begin{eqnarray}
{\cal M}^{(a)}&=&-\left({g\over\sqrt{2}}\right)^2{1\over f_\pi}
                      {1\over (p_1-p_3)^2-m_\rho^2}
                \left(1-{\delta\over 2m_\rho^2} (p_1-p_3)^2\right)
\nonumber\\
&&\times\Biggl[\biggl(\eta_1 p_4\cdot(p_1-p_3)-\eta_2p_2\cdot(p_1-p_3)\biggr)
             (p_{1\,\mu}+p_{3\,\mu})
\nonumber\\
&&\qquad\qquad
      -\biggl(\eta_1 p_4\cdot(p_1+p_3)-\eta_2p_2\cdot(p_1+p_3)\biggr)
             (p_{1\,\mu}-p_{3\,\mu})\Biggr]
       \epsilon^{\mu}_{a_1}(p_2),
\\[12pt]
{\cal M}^{(b)}&=&-\left({g\over\sqrt{2}}\right)^2{1\over f_\pi}
                      {1\over (p_1-p_4)^2-m_\rho^2}
                \left(1-{\delta\over 2m_\rho^2} (p_1-p_4)^2\right)
\nonumber\\
&&\times\Biggl[\biggl(\eta_1 p_3\cdot(p_1-p_4)-\eta_2p_2\cdot(p_1-p_4)\biggr)
             (p_{1\,\mu}+p_{4\,\mu})
\nonumber\\
&&\qquad\qquad
      -\biggl(\eta_1 p_3\cdot(p_1+p_4)-\eta_2p_2\cdot(p_1+p_4)\biggr)
             (p_{1\,\mu}-p_{4\,\mu})\Biggr]
       \epsilon^{\mu}_{a_1}(p_2),
\\[12pt]
{\cal M}^{(c)}_1&=& {1\over f_\pi^3} \Biggl[
 \biggl(2 d_1 f_\pi^2-d_2f_\pi^2 +2d_3 p_3\cdot p_4-2d_4 (p_4+p_3)
\cdot p_2\biggr) p^\mu_1
\nonumber\\
&&\qquad\qquad
-\biggl(d_2 f_\pi^2 +2d_3 p_1\cdot p_4-2d_4 p_1\cdot p_2\biggr) p^\mu_3
\nonumber\\
&& \qquad\qquad 
-\biggl(d_2 f_\pi^2 +2d_3 p_1\cdot p_3-2d_4 p_1\cdot p_2\biggr) p^\mu_4
\Biggr]\epsilon^{\mu}_{a_1}(p_2),
\\[12pt]
{\cal M}^{(c)}_2&=&{1\over f_\pi^3} \Biggl[
 \biggl(2d_2 f_\pi^2 -4d_3 p_3\cdot p_4+2d_4 (p_3+p_4)\cdot p_2\biggr) p^\mu_1
\nonumber\\
&&\qquad\qquad
-\biggl(2d_1 f_\pi^2 -2d_3 p_1\cdot p_4+2d_4 p_1\cdot p_2\biggr) p^\mu_3
\nonumber\\
&&\qquad\qquad  
-\biggl(2d_1 f_\pi^2 -2d_3 p_1\cdot p_3+2d_4 p_1\cdot p_2\biggr) p^\mu_4
\Biggr]\epsilon^{\mu}_{a_1}(p_2).
\end{eqnarray}

The invariant scattering matrix element is given by 
\begin{eqnarray}
{\cal M}(\pi^{+}+a_1^{-}\to\pi^{0}+\pi^{0})
&=&{\cal M}^{(a)}+{\cal M}^{(b)}+{\cal M}^{(c)}_1,
\nonumber\\
{\cal M}(\pi^{+}+a_1^{+}\to\pi^{+}+\pi^{+})
&=&{\cal M}^{(a)}+{\cal M}^{(b)}+{\cal M}^{(c)}_2.
\end{eqnarray}

\eject

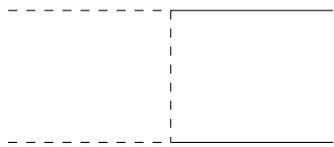
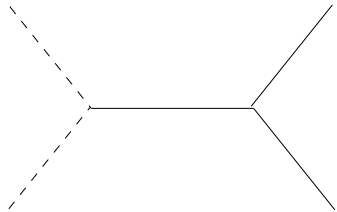
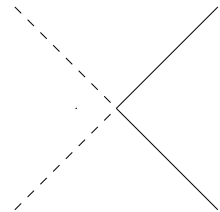
\begin{figure}
\caption{Diagrams for the reaction 
$\pi^{+}+\pi^{-}\to\rho^{+}+\rho^{-}$ and 
$\pi^{+}+\pi^{0}\to\rho^{+}+\rho^{0}$.
Dashed line and solid line indicate the pion and $\rho$ meson, respectively.}
\label{rrpp_a}
\end{figure} 

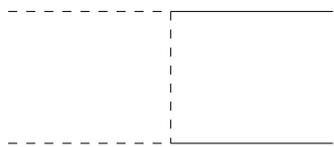
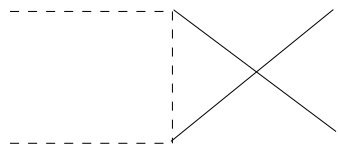
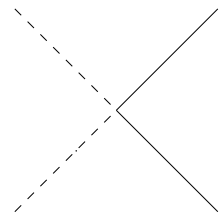
\begin{figure}
\caption{Same as Fig. 7 for the reaction
$\pi^{+}+\pi^{-}\to\rho^{0}+\rho^{0}$ and 
$\pi^{0}+\pi^{0}\to\rho^{+}+\rho^{-}$. }
\end{figure} 

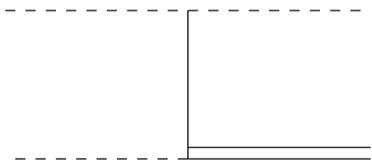
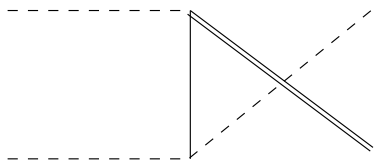
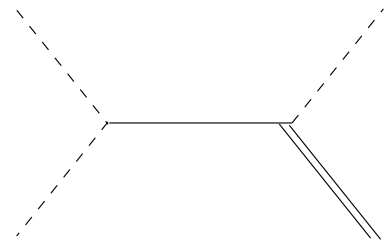
\begin{figure}
\caption{Diagrams for the reaction $\pi\omega\to\pi\pi$. Dashed line,
solid line and double solid line indicate the pion, $\rho$ meson and $\omega$
meson, respectively.}
\end{figure} 

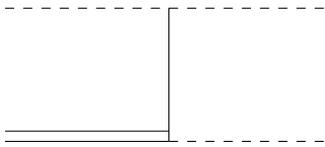
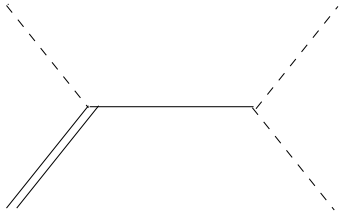
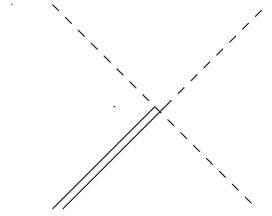
\begin{figure}
\caption{Diagrams for the reaction 
$\pi^{+}+a_1^{0}\to\pi^{0}+\pi^{+}$ and 
$\pi^{+}+a_1^{-}\to\pi^{+}+\pi^{-}$
Dashed line, solid line and double solid line indicate the pion, 
$\rho$ meson and $a_1$ meson, respectively.}
\label{papp_a}
\end{figure} 

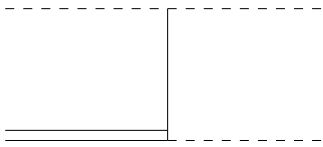
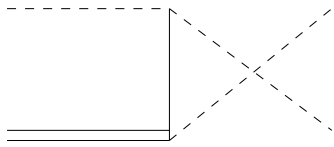
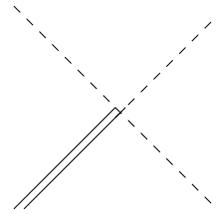
\begin{figure}
\caption{Same as Fig. 10 for the reaction
$\pi^{+}+a_1^{-}\to\pi^{0}+\pi^{0}$ and
$\pi^{+}+a_1^{+}\to\pi^{+}+\pi^{+}$.}
\end{figure} 

\newpage

\begin{references}


\bibitem{rhic} For a recent review see Quark Matter `95,
               Nucl. Phys. {\bf A590}, 1 (1995).

\bibitem{qgp} E. V. Shuryak, Phys. Report  {\bf 61}, 71 (1980).\\
              L. McLarren, Rev. Mod. Phys. {\bf 58}, 1001 (1986). 


\bibitem{shuryak} E. Shuryak, Phys. Rev. Lett. {\bf 68}, 3270 (1992).

\bibitem{wang} T. Bir\'o, E. Doorn, B. M\"uller, M. Thoma, and X. Wang,
               Phys. Rev. C {\bf 48}, 1275 (1993);\\
               X.-N. Wang, Nucl. Phys. {\bf A590}, 47c (1995).

\bibitem{geiger} K. Geiger and J. Kapusta, Phys. Rev. D {\bf 47}, 4905 (1993).


\bibitem{BM} J. Stachel and P. Braun-Munzinger, 
             Phys. Lett. B {\bf 216}, 1 (1989).

\bibitem{BM_AGS} P. Braun-Munzinger, J. Stachel, J. Wessels and N. Xu,
                 Phys. Lett. B {\bf 344}, 43 (1995).

\bibitem{BM_SPS} P. Braun-Munzinger, J. Stachel, J. Wessels and N. Xu,
                 Phys. Lett. B {\bf 365}, 1 (1996).

\bibitem{rafelski} J. Rafelski, Phys. Lett. B {\bf 262}, 333, (1991).\\
                   J. Letessier, A. Tounsi, and J. Rafelski, 
                   Phys. Lett. B {\bf 292}, 417 (1992).\\  
                   J. Letessier, J. Rafelski, and A. Tounsi,
                   Phys. Lett. B {\bf 328}, 499 (1994).

\bibitem{cleymans} N. J. Davidson, H. G. Miller, R. M. Quick and J. Cleymans,
                   Phys. Lett. B {\bf 255}, 105 (1991).

\bibitem{heinz} J. Sollfrank, M. Ga\'zdzicki, U. Heinz and J. Rafelski,
                Z. Phys. C {\bf 61}, 659 (1994).\\ 
                U. Heinz, Nucl Phys. {\bf A566}, 205c (1994). 


\bibitem{strange} J. Sollfrank and U. Heinz, 
                  in {\it Quark-Gluon Plasma 2}, edited by R. C. Hwa
                  (World Scientific, 1996).
 

\bibitem{chpt} J. Gasser and H. Leutwyler, 
               Phys. Lett. B {\bf 184}, 83 (1987).\\ 
               P. Gerber and H. Leutwyler, Nucl. Phys. {\bf B321}, 387 (1989).

\bibitem{thermal} J. L. Goity and H. Leutwyler, 
                  Phys. Lett. B {\bf 228}, 517 (1989).\\
                  Chungsik Song, Phys. Rev. D {\bf 48}, 1556 (1994).\\
                  K. Haglin and S. Pratt, Phys. Lett. B {\bf 328}, 255 (1994).

\bibitem{coll} K. Haglin, Nucl. Phys. {\bf A584}, 719 (1995).\\
               C. Song and C. M. Ko, Phys. Rev. C {\bf 53}, 2371 (1996). 


\bibitem{matsui} T. Matsui, B. Svetitsky and L. McLerran, 
                 Phys. Rev. D {\bf 34}, 783 (1986),{\it ibid}
                 Phys. Rev. D {\bf 34}, 2047 (1986).

\bibitem{goity} J. L. Goity, Phys. Lett. B {\bf 319}, 401 (1993).

\bibitem{formfactor} D. Lohse, K. W. Durso, K. Holinde and J. Speth,
                     Phys. Lett. B {\bf 234}, 235 (1990).
              

\bibitem{delta} G. F. Bertsch and S. Das Gupta,
                Phys. Rep. {\bf 160}, 189 (1988). 


\bibitem{bebie} H. Bebie, P. Gerber, J. L. Goity, and H. Leutwyler, 
                Nucl. Phys. {\bf B378}, 95 (1992).


\bibitem{mym} U.-G. Meissner, Phys. Rep. {\bf 161}, 213 (1988).

\bibitem{song} Chungsik Song, Phys. Rev. C {\bf 47}, 2861 (1993).

\bibitem{witten} E. Written, Nucl. Phys. {\bf B223}, 432 (1983).\\
                  \"O. Kaymakcalan, S. Rajeev, and J. Schechter,
                  Phys. Rev. D {\bf 30}, 594 (1984).

\end{references}
\end{document}